# Privacy Protection Cache Policy on Hybrid Main Memory

N.Y. Ahn and D.H. Lee

*Abstract*—We firstly suggest privacy protection cache policy applying the duty to delete personal information on a hybrid main memory system. This cache policy includes generating random data and overwriting the random data into the personal information. Proposed cache policy is more economical and effective regarding perfect deletion of data.

## I. Introduction

On, May 13, 2014, the Court of Justice of the European Union (CJEU) announced the historical decision in personal information protection, which is "the right to be forgotten" in the context of data processing on internet search engines [1], [2]. CJEU decided that the internet service providers (ISPs) of the search engines would be responsible for the processing of personal information in web pages by third parties [3]. Recently Sachiko Kanamori, Kanako Kawaguchi, and Hidema Tanaka introduced the scheme for the right to be forgotten using secret sharing and digital watermarking in social networking services (SNSs) [4]. And Hiroki Yamazawa, Kazuki Maeda, Tomoko Ogura Iwasaki and Ken Takeuchi at Chuo University proposed privacy-protection solid state storage (PP-SSS) system for internet data's "the right to be forgotten", in which data lifetime is specified without file system overhead [5]. The PP-SSS controls data lifetime using precision error correction code (ECC) and crush techniques. Naturally, the right to be forgotten in system memories should be considered. Especially, if system memories are configured to include a non-volatile memory [6], the internet providers ISPs should design the system memories to meet "the duty to delete" in the non-volatile memory, overwhelming "the right to be forgotten".

## II. Rerated Works

Recently, hybrid main memory systems include a central processing unit (CPU), a volatile memory such as a dynamic random access memory (DRAM) and a non-volatile memory (NVM) such as a 3D Xpoint memory, a NAND flash memory, a phase change memory (PCM), a spin transfer torque random access memory (STT-RAM), a ferro-magnetic RAM (FeRAM), etc referring to Fig. 1 [7], [8], [9], [10]. CPU may access caches of DRAM in processes. If a cache is not used by CPU for a predetermined time, the dirty cache is flushed to NVM according to a cache policy [11], [12], [13], [14]. That is, DRAM flushes the dirty cache to NVM. Then the dirty

N.Y. Ahn is a student at Graduate School for Information Management, Korea University, 145, Anam-ro, Seongbuk-gu, Seoul, Korea (email: humble@korea.ac.kr)

D.H. Lee is a professor at CIST and Graduate School for Information Security, Korea University (e-mial: donghlee@korea.ac.kr)

cache still remains in NVM. After flushing, the dirty cache may be updated by CPU. How to manage the dirty cache in NVM? In general, data of NVM are managed by the mapping table. The mapping table is used to translate logical addresses to physical addresses. If the dirty cache is personal information, we need to delete completely the personal information for the privacy protection.

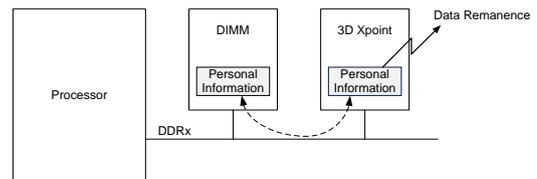

FIG.1 Hybrid Main Memory.

But, complete deletion of NVM cannot be achieved. This is because the physical deletion is too expensive in regard with time, power consumption, etc. Generally, the mapping table of NVM is only changed to CPU request, still existing the original data, that is personal information in NVM. In fact, Juniang Shu etc. studied data a remanence experiment on mobile devices: data cleaning, application uninstallation, factory reset [15]. At least 40% data remanence rate of the target deleted files still remains on mobile devices by 9 weeks. If the target deleted files (that is cache) are personal information, this situation is very serious. We have to strongly apply the duty to delete cache data on NVM. We are called to Privacy Protection Cache Policy on the hybrid main memory.

## III. Privacy Protection Cache Policy

We suggest that the original cache data be overwritten to random cache data in response to the privacy protection request of CPU. Herein the random cache data are internally generated in NVM on receiving the request from CPU. The generation schemes of random cache data change according to types of NVM. For example, if NVM is an over-writable memory, such as 3D-Xpoint, PRAM, MRAM, ReRAM, etc., the random cache data may be generated by a random number. Then NVM overwrites the generated random cache data into the corresponding cache data. On the other hand, if NVM is not over-writable memory, such as a NAND flash memory, the random cache data may be only generated in limited environments. According to the above Privacy Protection Cache Policy, we utilize random data to overwrite the original cache data. As a result, the dirty cache having personal information is changed into random cache data.

## IV. Privacy Protection Cache Policy on Hybrid Main Memory

We introduce the privacy protection mode of the hybrid main memory bellows. In the privacy protection mode, DIMM may receive overwrite request from the processor, and delete personal information by overwriting random data. Where the random data may be transferred to DIMM with the overwrite request. And, in the privacy protection, 3D-Xpoint memory may receive the deletion request for personal information, search valid/invalid personal information in response to the deletion request, delete the searched personal information by overwriting random data, and verify the personal information's existence. If the personal information is not exist in the verification, 3D-Xpoint memory issues the deletion completion and transfers the deletion completion to the processor.

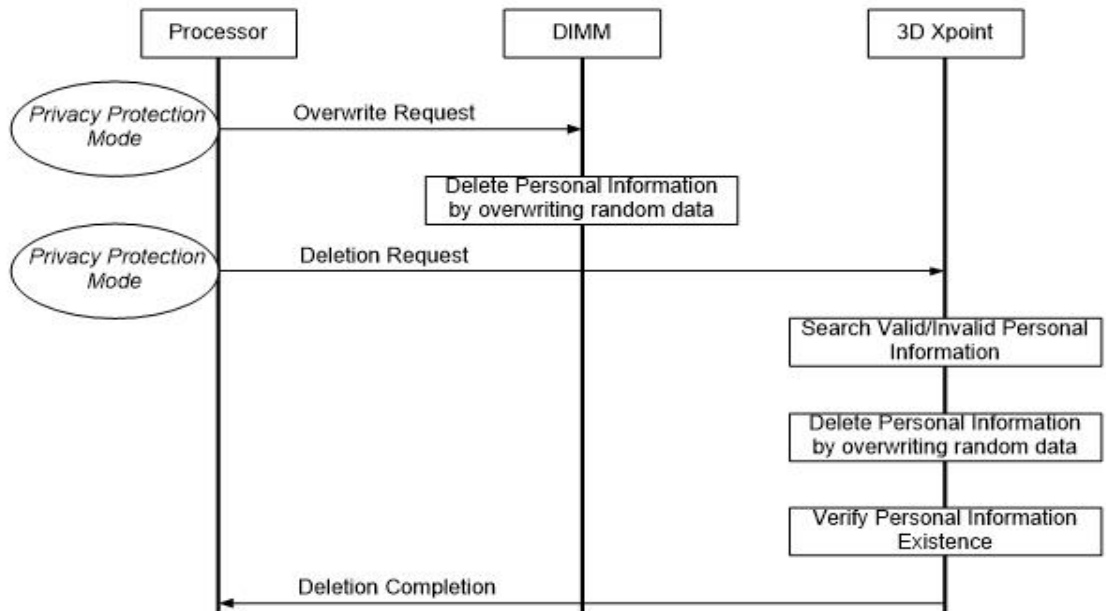

FIG. 2. Hybrid Main Memory in Privacy Protection Mode.

### A. Random Data Generation

The 3D-Xpoint memory may generate random data in response to the deletion request from the processor. The random data may be generated by the random number generator in the 3D-Xpoint memory.

### B. Fully/Partially Overwriting

Firstly, the personal information is fully overwritten according to the random data. Secondly, the personal information is partially overwritten according to the random data. Accordingly, types of the random data are determined by the above overwriting degree about the personal information, referring to FIG. 3.

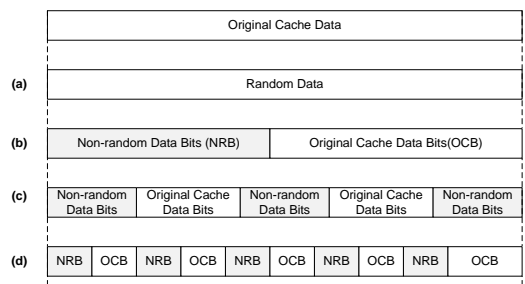

FIG. 3 Non-random Data Types.

## V. CONCLUSION

We introduced the privacy protection cache policy on the hybrid main memory. The privacy protection cache policy includes overwriting personal information by random data in DIMM/NVM. Proposed cache policy is more economical and effective for the privacy protection. In future, we need to study new cache architectures to meet "the duty to delete" in main memory systems.

ACKNOWLEDGMENT

N.Y. Ahn thanks to my God and my academic advisor, Prof. Lee for giving an opportunity to study Ph.D. course in Information Security.

REFERENCES

[1] Judgment of the Court (Grand Chamber),13 May 2014; http://curia.europa.eu/juris/liste.jsf?num=C-131/12
[2] B. Kampmark, "To Find or be Forgotten: Global Tensions on the Right to Erasure and Internet Governance," Journal of Global Faultlines, Vol.2, No.2, 2015.
[3] K. O'Hara, "The Right to Be Forgotten: The Good, the Bad, the Ugly," IEEE Computer Society, 2015.
[4] S. Kanamori, K. Kawaguchi, and H. Tanaka, "Study on a Scheme for the Right to Be Forgotten," ISITA2014, Melbourne, Australia, October 26-29, 2014.
[5] H. Hiroki Yamazawa, K. Maeda, T. O. Iwasaki and K. Takeuchi, "Privacy-Protection SSD with Precision ECC and Crush Techniques for 15.5× Improved Data-Lifetime Control," %1 2016 IEEE 8th International Memory Workshop (IMW) , 2016.
[6] 3D Xpoint, Available: https://en.wikipedia.org/wiki/3D_XPoint.
[7] J. Wang and B. Wang, "A Hybrid Main Memory Applied in Virtualization Environments," 2016 First IEEE International Conference on Computer Communication and the Internet, 2016.
[8] S. Bock, B. R. Childers, R. Melhem and D. Mossé, "Concurrent Migration of Multiple Pages in software-managed hybrid main memory," %1 2016 IEEE 34th International Conference on Computer Design (ICCD) , 2016.
[9] H. A. Khouzani, C. Yang and F. S. Hosseini, "Segment and Conflict Aware Page Allocation and Migration in DRAM-PCM Hybrid Main Memory," IEEE Transactions on Computer-Aided Design of Integrated Circuits and Systems, 2016.
[10] D. Kim, S. Yoo and S. Lee, "Hybrid Main Memory for High Bandwidth Multi-Core System," IEEE Transactions on Multi-Scale Computing Systems, Vol. 1, No: 3, pp. 138-149, 2015.
[11] A. Magdy, R. Alghamdi and M. F. Mokbel, "On main-memory flushing in microblogs data management systems," 2016 IEEE 32nd International Conference on Data Engineering (ICDE) , 2016.
[12] E. Lee, H. Kang, H. Bahn and K. G. Shin, "Eliminating Periodic Flush verhead of File I/O with Non-Volatile Buffer Cache," IEEE Transactions on Computers, Vol. 65, No: 4, pp. 1145 - 1157, 2016.
[13] Z. Fan, D. H. C. Du and D. Voigt, "H-ARC: A non-volatile memory based cache policy for solid state drives," 2014 30th Symposium on Mass Storage Systems and Technologies (MSST), 2014.
[14] M. Tarihi, H. Asadi and H. Haghdoost, "A Hybrid Non-Volatile Cache Design for Solid-State Drives Using Comprehensive I/O Characterization," IEEE Transactions on Computers, Vol 65, No: 6, 2016.
[15] J. Shu, Y. Zhang, J. Li, B. Li, and D. Gu, "Why Data Deletion Fails? A Study on Deletion Flaws and Data Remanence in Android Systems," ACM Transactions on Embedded Computing Systems, Vol. 16, No. 2, Article 61, Publication date: January 2017.